\documentclass[iop]{emulateapj}
\usepackage{amsmath}
\usepackage{natbib}

\newcommand{\be}[1]{\begin{equation} \label{eq:#1}}
\newcommand{\ee}{\end{equation}}

\newcommand{\HeI}{\ion{He}{1}}
\newcommand{\HeII}{\ion{He}{2}}
\newcommand{\SiI}{\ion{Si}{1}}

\newcommand{\pref}{\protect\ref}
\newcommand{\solrad}{\ifmmode{R}_{\rm S}\else${R}_{\rm S}$\fi}

\newcommand{\en}{\ifmmode{\varepsilon}_{\rm eV}\else${\varepsilon}_{\rm eV}$\fi}
\newcommand{\enk}{\ifmmode{\varepsilon}_{\rm keV}\else${\varepsilon}_{\rm keV}$\fi}
\newcommand{\eno}{\ifmmode{\varepsilon}\else${\varepsilon}$\fi}

\newcommand{\solmas}{\ifmmode{M}_{\rm S}\else${M}_{\rm S}$\fi}

\newcommand{\ctn}{\ifmmode\kappa\else$\kappa$\fi}

\newcommand{\term}[2]{\mbox{$\,^{#1}{\rm #2}$}}
\def\term#1 #2/{\mbox{$\,^{#1}{\rm #2}$}}

\newcommand{\hemults}{\HeI{} 1083~nm multiplet}
\newcommand{\hemult}{\HeI{} 1083~nm multiplet ($1s2s~^3\!S_1 - 1s2p~^3\!P^o_{2,1,0}$)}

\newcommand\lta { \mathrel {\hbox to 0pt {\lower 3.7pt \hbox{$\sim$}
      \hss} \raise 1.7pt \hbox{$<$}}}
\newcommand\gta { \mathrel {\hbox to 0pt {\lower 3.7pt \hbox{$\sim$}
      \hss} \raise 1.7pt \hbox{$>$}}}

\usepackage[normalem]{ulem}
\newcommand\revise[1]{{#1}}
\newcommand\delete[1]{{{#1}}}

\newcommand{\philemail}{judge@ucar.edu}

\newcommand{\luciaemail}{lucia.kleint@fhnw.ch}
\newcommand{\albertoemail}{asainz@ucar.edu}

\newcommand{\figrhessi}{
\begin{figure*}[] 
\epsscale{1.}
\plotone{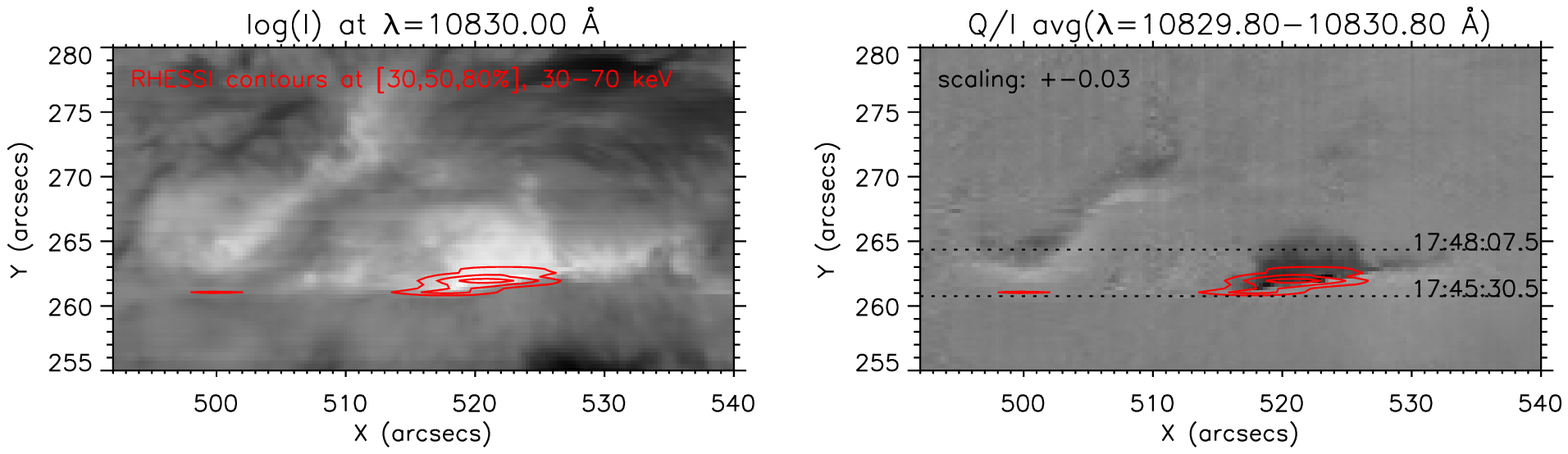}  
\caption{\label{fig:rhessi} Context maps of Stokes $I$ and $Q$ of the
  \hemults{} are shown, each row in each image corresponding to
  one integration.  
  The
  horizontal dashed lines span positions of the FIRS slit between the
  times shown, as it scanned from S to N on the plot.  HXR RHESSI
  contours at [30,50,80]\% of the peak flare intensity are plotted on
  top of the He intensity $I$ (left panel) and linear polarization $Q$
  (right panel).  The strongest HXR emission coincides with the
  strongest polarization, but significant polarization is seen outside areas
  detected by RHESSI.  }
\end{figure*}
}

\newcommand{\figseeing}{
\begin{figure*}[] 
\epsscale{.8}
\plotone{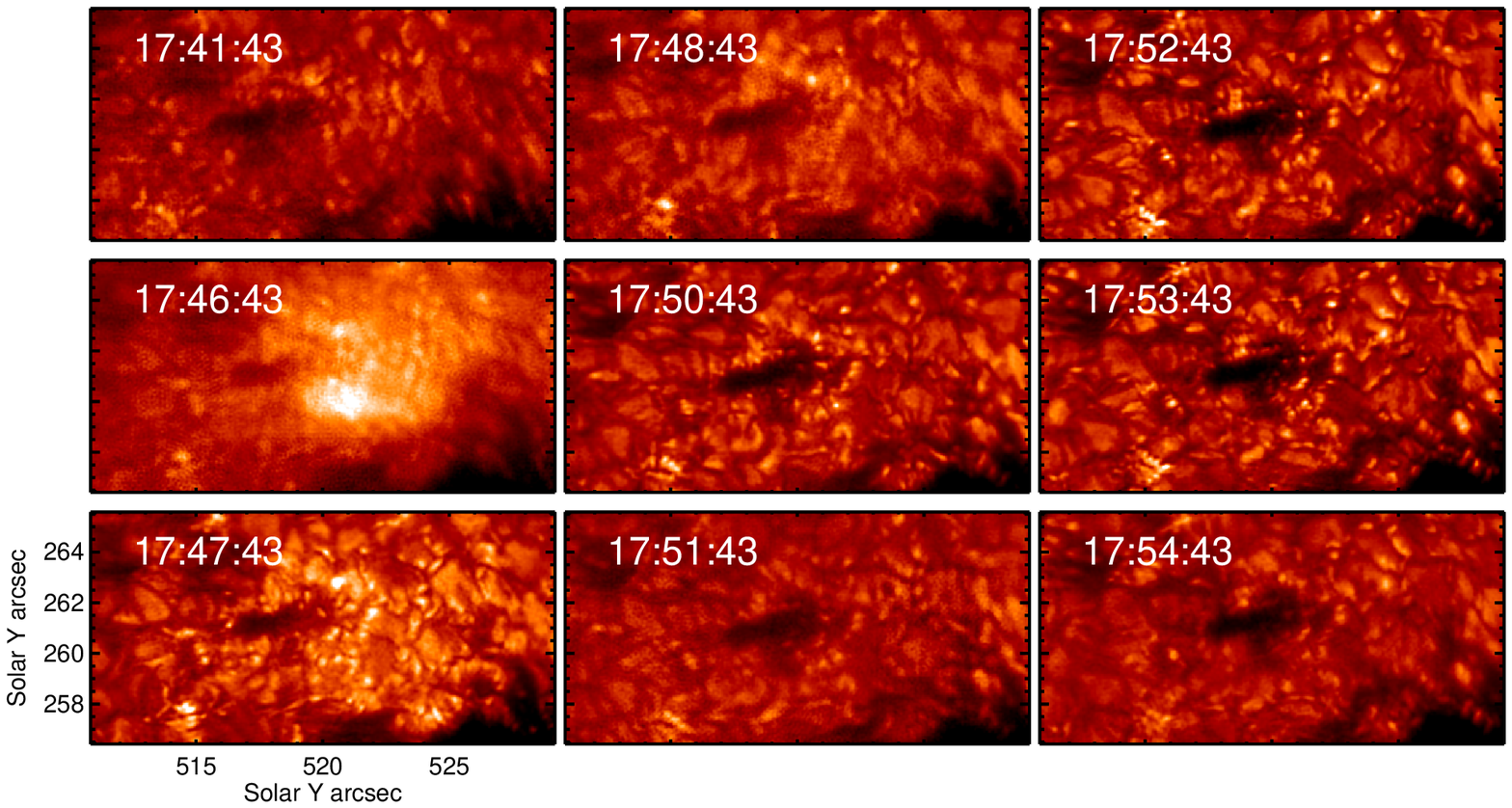}  
\caption{\label{fig:seeing} Speckle-reconstructed G-band images
  obtained with a CCD camera at the DST are shown, highlighting the
  changing quality of the atmospheric seeing during the impulsive
  phase of the flare ($\approx $ 17:45 to 17:48 UT).  The ``white
  light'' flare is seen in the 17:46:43 panel.  }
\end{figure*}
}

\newcommand{\figifit}{
\begin{figure}[] 
\epsscale{1.25}
\plotone{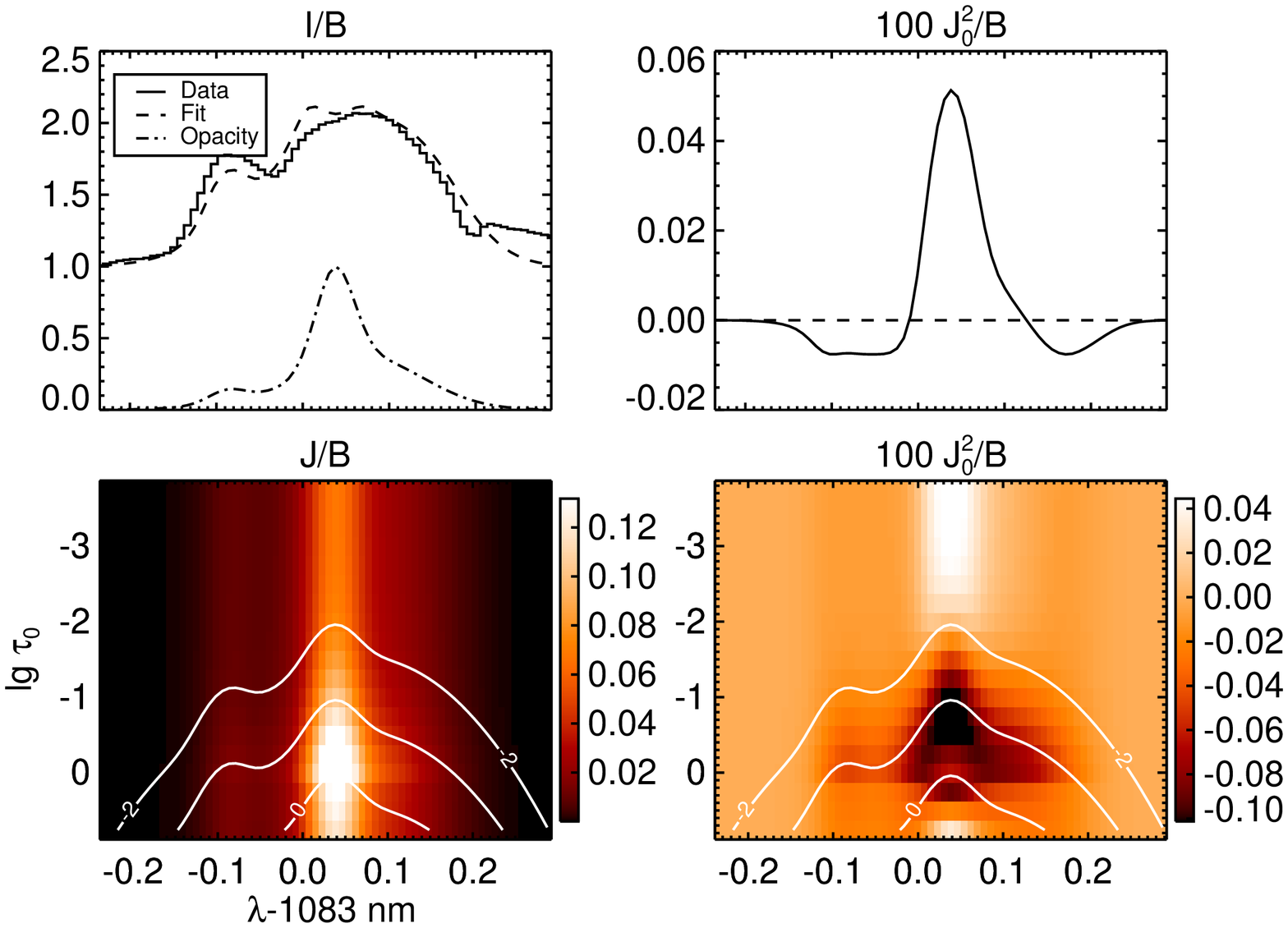}  
\caption{\label{fig:ifit} Results of non-LTE radiative transfer
  calculations in a slab are shown in which the slab parameters were
  optimized to fit the observed intensities (top left panel) from
  position $Y=41\arcsec$ from Figure \pref{fig:comparef}, part of the
  bright flare ribbon.  The model slab has an optical depth at the
  line core of 15, a source function with a thermal contribution of
  $\epsilon B$ where $\epsilon = 0.011$ and a local emission term with
  $B=10\times$ the 5700K photospheric Planck function, corresponding
  to a temperature of $2\times10^4$K. The upper right panel shows the
  radiation anisotropy as a function of wavelength emerging from the
  top of the slab, the lower panels expand the mean intensity and
  anisotropy as a function both of wavelength and optical depth in the
  slab (the solar photosphere lies just beneath the bottom of these
  plots, represented by upcoming radiation which is a Planck function
  at 5700K, independent of slant angle). Contours mark logarithmic
  values of monochromatic optical depth.}
\end{figure}
}

\newcommand{\td}{
\begin{figure}[] 
\epsscale{1.1}
\vskip 10pt
\plotone{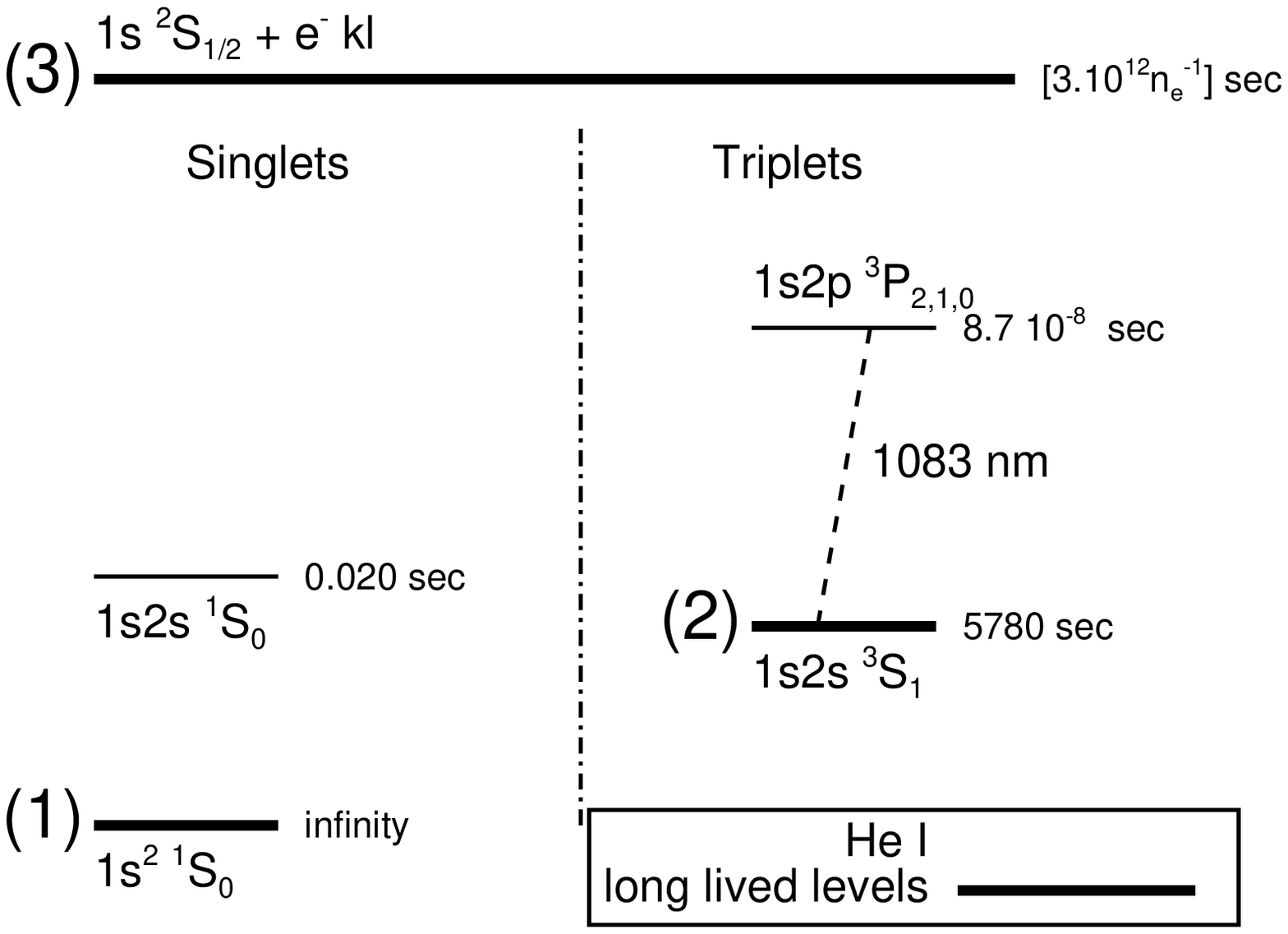}  
\caption{\label{fig:td} Energy levels of \HeI{} are shown, including
  the long-lived levels and the levels between which the 1083 nm transitions
  occur.  The lifetimes of the levels are listed along with the
  leading terms in the wavefunctions appropriate for each level.  The
  lifetime of the \HeII{} ground level is infinite {\em in vacuo}, the
  lifetime listed is the recombination time for a He$^+$ ion embedded
  in a plasma with electron density $n_e$ for temperatures near 10$^4$
  K.  }
\vskip 10pt
\end{figure}
}

\newcommand{\figcomparef}{
\begin{figure}[] 
\epsscale{1.2}
\plotone{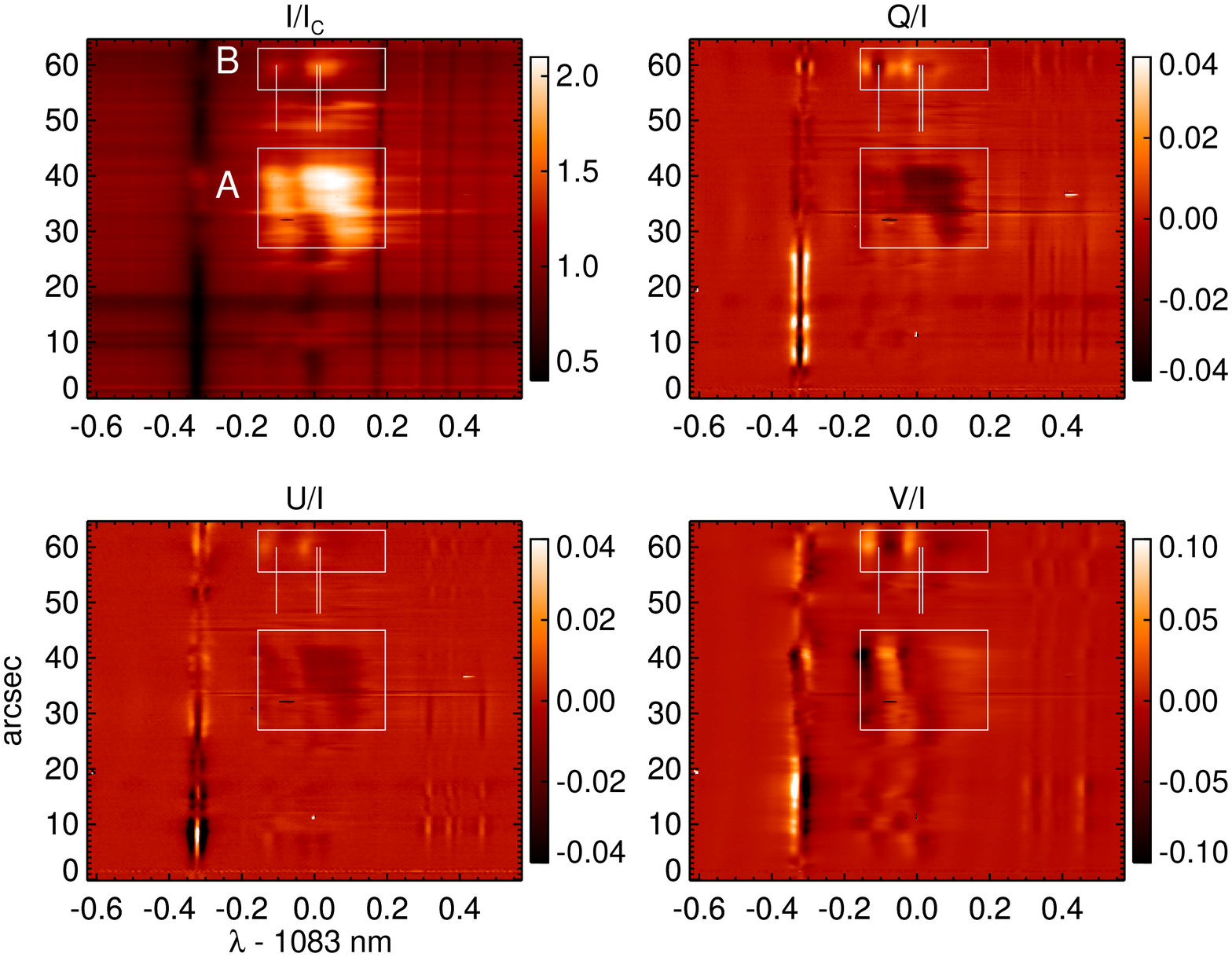}  
\caption{\label{fig:comparef} Stokes profiles are shown for one slit
  position of FIRS acquired over the 13 seconds after 17:46:42 UT.
  Wavelength increases from left to right (each pixel is 0.0039 nm, 39 m\AA),
  and the ordinate measures position from W to E along the projected
  slit in units of $0\farcs3$ (the zero point is arbitrary).  
  The three vertical lines show the three transitions of
  \ion{He}{1}, the two rightmost are blended.  The boxed regions
  highlight the bright footpoint that also shows the unusual profiles
  of $Q$ and $U$.  ``A'' and ``B'' mark footpoint emission areas, for
  comparison with later figures.   }
\end{figure}
}

\newcommand{\figmore}{
\begin{figure}[] 
\epsscale{1.0}
\plotone{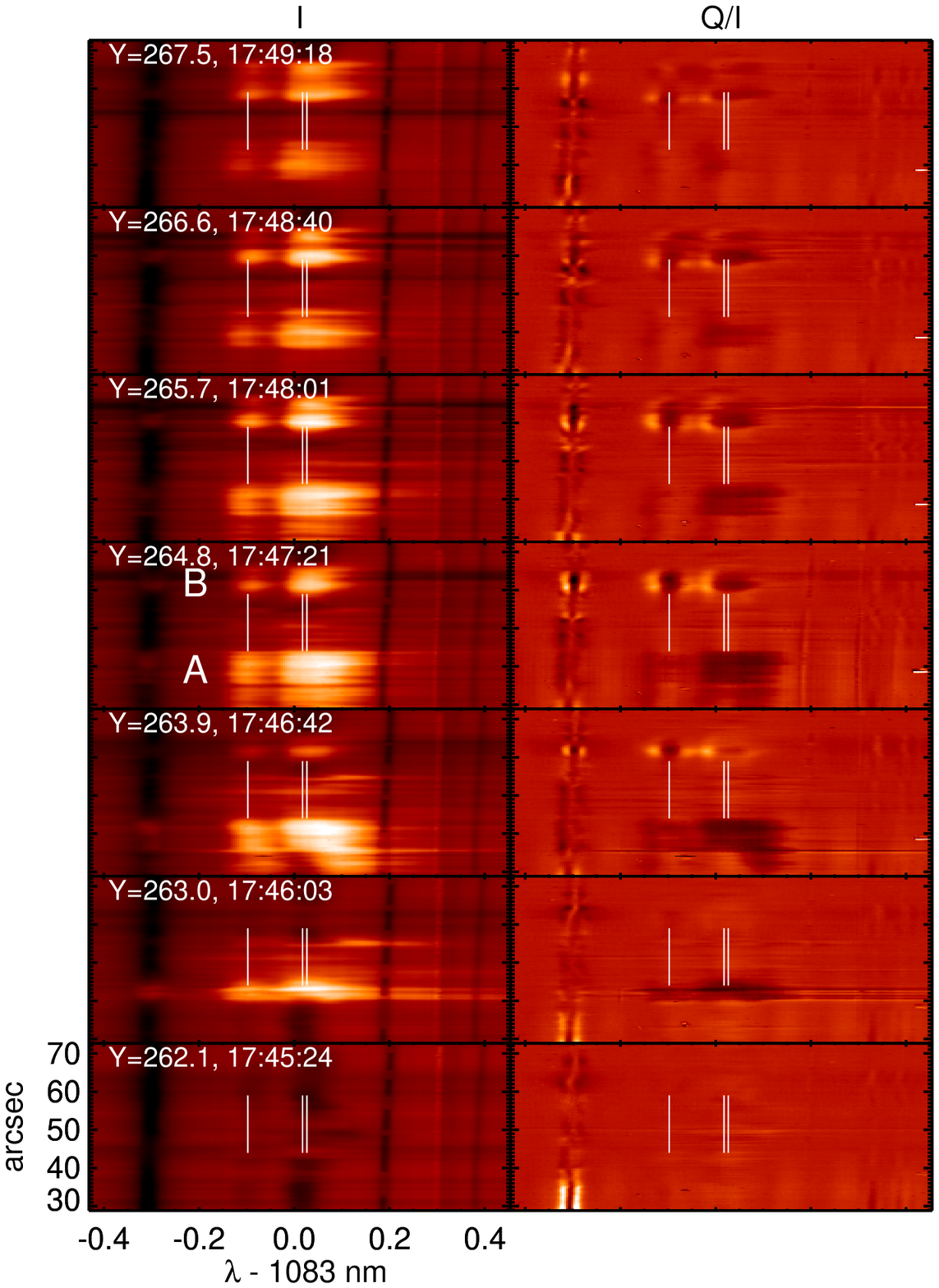}  
\caption{\label{fig:more} Stokes $I$ (left) and $Q$ (right) profiles
  are shown for seven position of the FIRS slit on the Sun, each
  integrated during the 13 seconds after the times shown, which are
  separated by 40 s.  The solar $Y$ position of the slit is listed
  along with the times, which can be seen in Figure~\pref{fig:rhessi}.
  The scales and intensities of the images are byte scaled to the same
  ranges shown in Figure~\pref{fig:comparef}.  }
\end{figure}
}

\newcommand{\lineplot}{
\begin{figure*}[] 
\epsscale{1.0}
\plotone{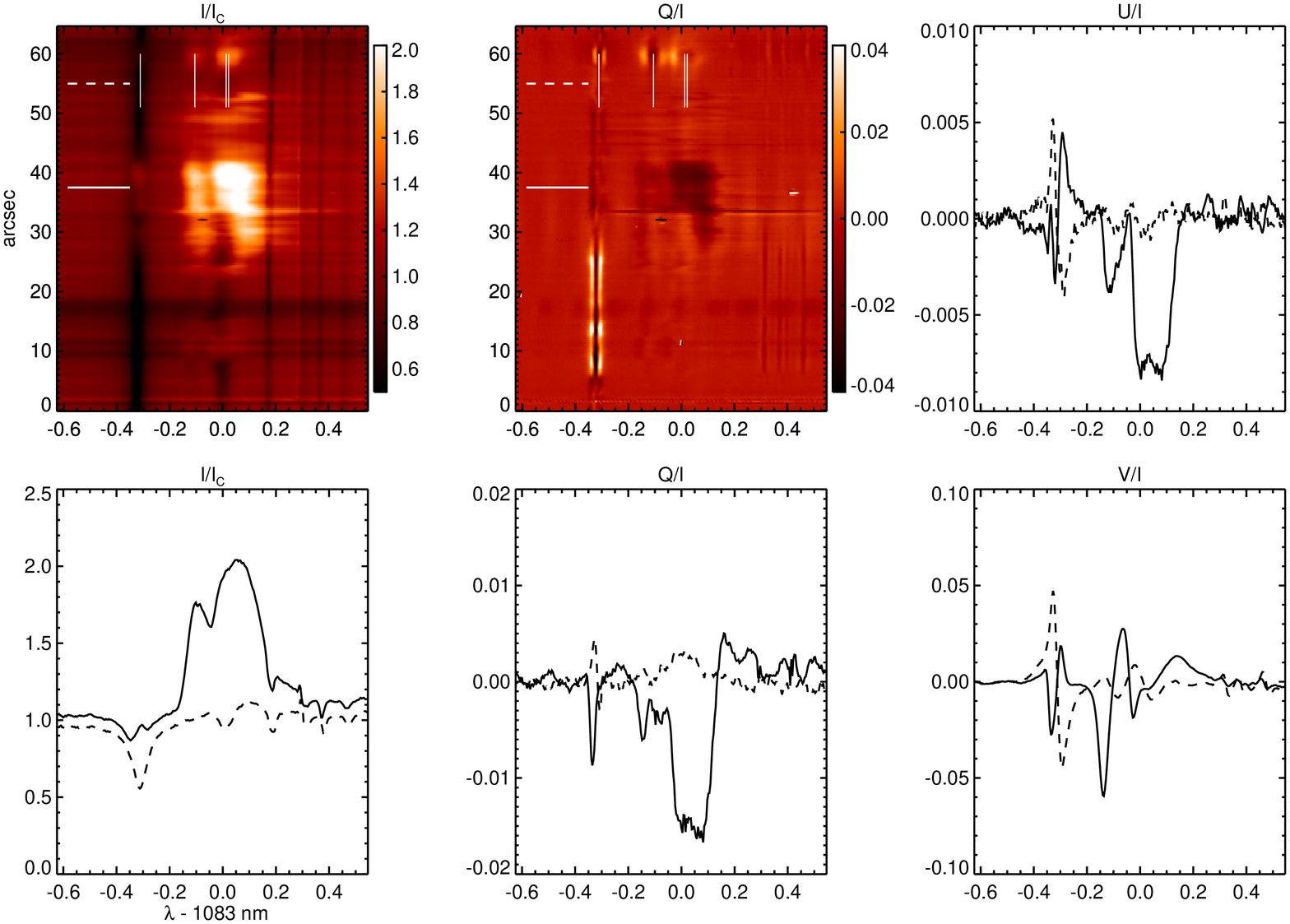}  
\caption{\label{fig:lineplot}
Line plots of 
Stokes $I$ and $Q,U,V$ profiles are shown along with $I,Q$ 
images shown earlier.  All plots are divided by the 
intensity.  The 1082.7 nm line of \SiI{} is marked as a vertical
line in the images,  along with 
the three lines of the \hemults{} to the right.   Dashed lines show
the Stokes profiles for $Y=5\arcsec$, and solid lines profiles for 
position $Y=38\arcsec$ 
during the bright impulsive phase.   
}
\end{figure*}
}

\newcommand{\percival}{
\begin{figure*}[] 
\epsscale{0.8}
\plotone{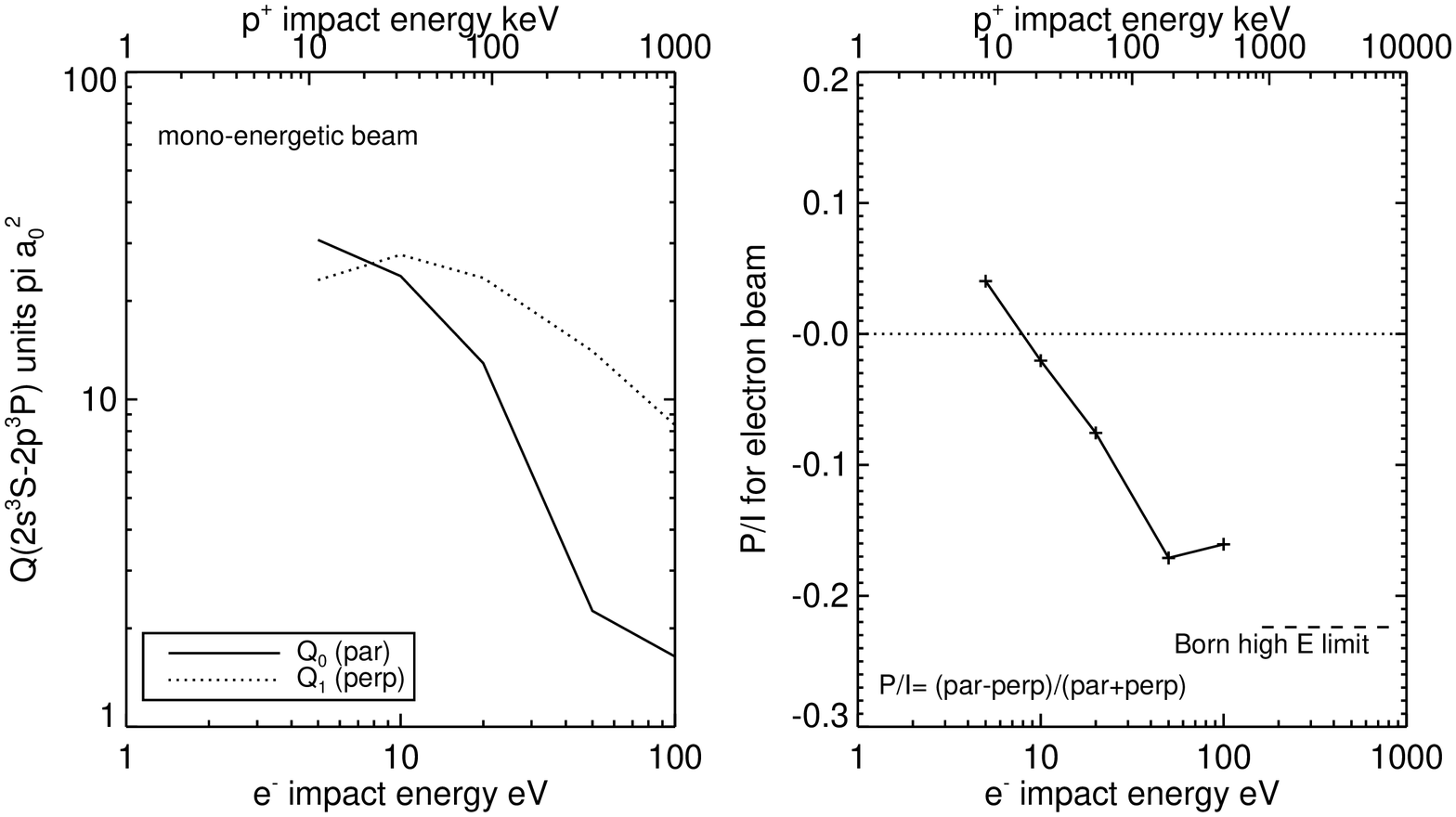}  
\caption{\label{fig:percival}
The left panel shows cross sections computed as a function of energy
for the excitation of parallel and perpendicular components of the 
\hemults{}.  (The reference direction is the direction of
the peak of the particle distribution function, a beam in this case). 
The sum of the cross sections gives the usual
total cross section for excitation from the long-lived 
$1s2s~^3S_0$ level in \HeI.  The right panel shows the energy dependence of the 
ratio of these components, related to the net linear polarization and its sign.  
}
\end{figure*}
}

\shortauthors{Judge et al.}
\shorttitle{Helium line polarization}

\slugcomment{}

\begin{document}

%
%

\title{On helium 1083 nm line polarization during
  the impulsive phase of an X1 flare}
\author{  
Philip G. Judge}
\affil{High Altitude Observatory,
       National Center for Atmospheric Research\footnote{The National %
       Center for Atmospheric Research is sponsored by the %
       National Science Foundation},\\
       P.O.~Box 3000, Boulder CO~80307-3000, USA; \philemail}


\author{Lucia Kleint}
\affil{ Institute of 4D Technologies, University of
  Applied Sciences and Arts Northwestern Switzerland,\\ 5210 Windisch,
  Switzerland; \luciaemail}

\author{Alberto Sainz Dalda} 
\affil{High Altitude Observatory,
       National Center for Atmospheric Research,\\  P.O.~Box 3000, Boulder CO~80307-3000, USA; 
\albertoemail} 

%
%

\begin{abstract}
\revise{
We analyze spectropolarimetric data of the \hemult{} during the X1
flare SOL2014-03-29T17:48, obtained with the Facility Infrared
Spectrometer (FIRS) at the Dunn Solar Telescope.  While scanning
active region NOAA 12017, the FIRS slit crossed a flare ribbon during
the impulsive phase, when the helium line intensities
turned into emission at $\lta$ twice the continuum intensity. 
Their linear polarization profiles are of the same sign across the
multiplet including 1082.9 nm, 
intensity-like,  at $\lta5$\% of the continuum intensity.
Weaker Zeeman-induced linear polarization is also observed.
Only the strongest linear polarization coincides  with hard
X-ray (HXR) emission at 30-70 keV observed by the \textit{Reuven
  Ramaty High Energy Solar Spectroscope Imager}. The 
polarization is generally 
more extended and lasts longer than the HXR
emission.
The upper $J=0$ level of the 1082.9~nm component is unpolarizable,
thus lower level polarization is the culprit.  We make non-LTE
radiative transfer calculations in thermal slabs optimized to fit only
intensities.  The linear
polarizations are naturally reproduced, through a systematic change of sign with
wavelength of the radiation anisotropy
when slab optical depths of the 1082.9
component are $\lta 1$.  Collisions with beams of
particles are neither needed nor can they produce the same sign of
polarization of the 1082.9 and 1083.0 nm components.  The
\ion{He}{1} line polarization merely requires heating sufficient to
produce slabs of the required thickness.  Widely different
polarizations of H$\alpha$, reported previously, are explained by
different radiative anisotropies arising from slabs of different
optical depths.}

\end{abstract}

\keywords{Sun: atmosphere - Sun: chromosphere - Sun:flares - Sun:protons}

%
%
\section{Introduction}

The physics of solar flares is still not understood in spite of over
80 years of quantitative research, arguably beginning with the work of
\citet{Newton1930} and \citet{Hale1931}.  As reviewed by
\citet{benz2008rev,Fletcher+others2011,Holman+others2011}, flares are
believed to originate from the slow buildup of magnetic energy in the
corona, followed by sudden or ``impulsive'' energy release, most
likely via reconnection \citep{Carmichael1964,Sturrock1966,
  Hirayama1974,Kopp+Pneuman1976}. But there remain elementary
unanswered questions. For example, the origin of broad-band (``white
light'') emission, observed for over 150 years \citep{Carrington1859},
is still debated \citep{Kerr+Fletcher2014,Heinzel+Kleint2014}.

\figrhessi

Given our rudimentary knowledge of flare physics, it is important to
obtain critical observations.  Here we analyze unusual polarimetric
data of chromospheric \hemult{} observed during the impulsive phase of
the X1 flare SOL2014-03-29T17:48.  We study the origin of linear
polarization $P=\sqrt{Q^2+U^2}$ (with $Q$ and $U$ the usual Stokes 
parameters) in this multiplet which have a peak magnitude $|P|$ a
few percent of intensity $I$.  \citet{Kuckein+others2015a,
  Kuckein+others2015b} report similar data for an M3.2 class flare
with $|P|/I$ an order of magnitude smaller, but with no analysis of
the origin of the polarization. 

Figure \pref{fig:rhessi} shows images of the active region
reconstructed from the FIRS scan during which the flare occurred,
beginning at 17:40:06 UT on 29 March 2014, ending at 18:01:39 UT with
Stokes $I$ at 1083.0 nm and $Q$ data, averaged over 0.1 nm. 
Each row of the image
corresponds to one slit spectrum in FIRS, the lowest row shows the
first spectrum, the next shows the spectrum acquired 13 seconds later,
and so forth.  Also shown are contours from the \textit{Reuven Ramaty
  High Energy Solar Spectroscope Imager}
\citep[RHESSI,][]{Lin+others2002}.  RHESSI records hard X-ray (HXR)
emission which in flares is generally caused by accelerated electrons.
Although the maximum polarization signal correlates with the maximum
in HXR counts, lower levels of linear polarization are clearly visible
where no HXR emission exists (e.g., near X=530\arcsec, Y=263\arcsec).  These
relationships will be discussed below, keeping an open mind as to the
physical relationship between these measurements.


The X1 flare of 29 March 2014, like many flares, exhibits hard X-ray
emission whose spectral properties in the standard model 
require supra-thermal electrons
propagating down from the corona
\citep[e.g.,][]{Brown1971,Fletcher+others2007}.  Such anisotropic
distributions (``beams'') of particles can carry significant energy
and momentum down to the solar atmosphere, from which white
light emission seems to originate. 
Anisotropic particle distributions can induce ``atomic
polarization'' which in turn generates 
spectral line polarization \citep{Percival+Seaton1958}.
Observations in the Balmer-$\alpha$ line of hydrogen at 656.3 nm during flares
have been made to seek evidence for such 
 anisotropies 
\citep{Henoux+Chambe1990,Henoux+others1990,Hanaoka2003,Bianda+others2005}.
But as yet there is no consensus on properties of H$\alpha$
polarization from flare ribbons. In part this is because ribbons are
difficult to observe with a spectrograph 
because flares are not predictable.
H$\alpha$ sometimes has 
$P/I$ at a level of a 5 percent \citep{Hanaoka2003}.
At other times, and with different instruments, it appears to show no
polarization, with $P/I < 0.5\%$ \citep{Bianda+others2005}.  Different
flares behave differently. The polarized spectrum can depend on the
phase of flare evolution, on the distance of the flare from disk
center, on the strength of the flare, the configuration of the
magnetic field, among the possibilities.  Polarimetric observations
themselves also use diverse 
spectral lines, different instruments 
(gratings and/or filters), under different 
 angular resolution, seeing and cross-talk conditions.
Thus the diversity {\revise of} H$\alpha$ polarized light
from flares is not surprising, but it is large and the
origin of the differences are not known.

Atomic polarization is generated also under conditions of asymmetry in
the radiation fields \citep[e.g.,][]{Casini+Landi2008}.  In
the present work we examine the helium data with the aim of
identifying the origin of the polarization.  The \hemults{} has
advantages over H$\alpha$: fine structure is partly split in solar
spectra -- it consists of just three lines (two blended at 1083.025
and 1083.034 nm, and the 1082.909 nm
line\footnote{\url{http://www.nist.gov/pml/data/asd.cfm}}), there is
no additional complexity from hyperfine structure, and, importantly,
the $J=0$ upper level of the 1082.909~nm transition is intrinsically
unpolarizable.  In Figure~\pref{fig:td} we show this
multiplet's levels along with the long-lived levels of helium, in a
highly simplified term diagram.

\td

%
%

\section{Analysis}

\subsection{Observations}

We obtained spectropolarimetric data with the Facility Infrared
Spectrometer \cite[``FIRS''][]{Jaeggli2011} at the Dunn Solar
Telescope (``DST'') of the National Solar Observatory in Sunspot, New
Mexico.  In our recent paper \citep[][henceforth
  ``Paper~I'']{Judge+others2014}, we reported observations of the
footpoint of the X1 flare SOL2014-03-29T17:48 in active region NOAA
12017.  In Paper I we analyzed the continuum and line of \ion{Si}{1}
at 1082.7 nm with the aim of explaining an associated sunquake.  Here
we analyze new reductions of the same scans but focusing on the
\hemults{}.

\figcomparef

\revise{
Figure~\pref{fig:comparef} shows data from one
slit position obtained in the flare ribbon 
during the impulsive phase,
from one integration from 17:46:42 {--} 17:46:55 UT.  Wavelength is
along the abscissa, position along the slit on the ordinate.  The peak of
the \ion{He}{1} emission arises from $X=520\arcsec$, $Y=264\arcsec$,
close to the brightest part of the flare seen at other infrared
wavelengths (Paper~I).  The three vertical lines show the approximate
rest positions of the \ion{He}{1} 1083 nm multiplet.  
The strong
absorption line of \ion{Si}{1} lies along $X=-0.3$ nm, the
line core also goes into emission near $Y=37$.
The \ion{Si}{1} line exhibits Stokes $Q,U$ and $V$ profiles which are
proportional to second and first derivatives with wavelength of Stokes
$I$ respectively. These profiles arise from the Zeeman effect
\citep[e.g.][]{Jefferies+Lites+Skumanich1989, Lites2000}.  
}

\delete{Figure \pref{fig:comparef} shows Stokes
profiles as a function of wavelength and position along the FIRS slit,
during the flare's impulsive phase. }

Unlike
the \ion{Si}{1} line, the \ion{He}{1} impulsive phase spectra show
signatures of strong atomic polarization, with Stokes $Q$ and $U$
profiles giving a polarization $P/I$ which is several
percent of the intensity, $I$, during the flare.  The impulsive phase linear
polarization in the \hemults{} has profiles similar but not identical
to Stokes $I$, with peak signal-to-noise ratios of at least 20.
The ``reference direction for linear polarization'' was E-W on the solar disk
(i.e. positive $Q$ goes from E to W).  

\subsection{Data reduction}

We re-reduced the FIRS data reported in Paper~I using the official NSO
software package written by C. Beck
(\url{http://nsosp.nso.edu/dst-pipelines}).  The new reductions
process the data minimally to optimize the purity of the
polarization signals. In particular, the 
flat fielding algorithm makes no assumptions
concerning the solar line profiles of interest, and the polarimetric
calibration includes a correction of $I \rightarrow [Q,U,V]$
cross-talk using continuum pixels which we assume, to the
sensitivities we achieve ($P/I \approx 10^{-3}$), are
polarization-free.

In Paper~I we measured noise levels of $8\times10^{-4}I_C$, where
$I_C$ is the median pre-flare continuum intensity.  However, the
reduced polarization data contain obvious optical fringes, the largest
of which are seen in Stokes $V$, at the level of $2\times10^{-3}I_C$.
These fringes are not insignificant when discussing the \HeI{} $Q,U,V$
spectra.  We tried and failed to remove fringes using the PCA-based
algorithm of \citet{Casini+Judge+Schad2012}. We
simply could not make the fringes belong only to one set of components
and the solar data to another, orthogonal set.  Instead, for wavelengths
close to the \hemults{}, we noted that almost no polarization was
detectable in the first and last slit positions of the scan during
which the flare occurred.  By assuming that these data only contain
fringes and other slowly varying artifacts, we made a linear
interpolation in time between the first and last scans for $Q,U$, and
subtracted the interpolated data from the $Q,U$ spectra.  While not
perfect, the procedure removed the bulk of the fringe pattern (compare
Figure~\pref{fig:comparef} with Figure~2 of Paper~I.)  
However there remain artifacts visible
as diffuse vertical stripes in the center of the Stokes $Q$
panel of Figure~\pref{fig:comparef}.  This is unfortunate since the
linear polarization of the 1082.9 and 1083.0 nm components of the
\hemults{} is expected to show both positive and negative $Q$ and $U$
(section~\pref{sec:disc}).

No polarization is visible in the telluric line of
H$_2$O lying along the column near $1083.2$~nm in Figure
\pref{fig:comparef}.  The upper limit to any
polarization at these wavelengths is below $10^{-3}I_C$, far less than the 
$\gta 10^{-2}I_C$ values of $Q$ and $U$ found during the flare scan.
Thus, at the wavelengths of the telluric line, errors in the
polarization calibration are significantly smaller than the $Q,U$
signals we are trying to obtain from these data.  However, this
encouraging analysis cannot na\"ively be applied to every wavelength in the
spectrum in the presence of seeing- or evolutionary-driven crosstalk
for  reasons given
in Appendix \pref{sec:timevar}.

During the impulsive phase, the \ion{He}{1} lines also show hints of
Zeeman-induced patterns near $Y=60\arcsec$ (box ``B'').  But the \ion{He}{1}
polarization is strikingly different in the region of strong flare
emission in the region $25\arcsec < Y < 41\arcsec$ where $Q$ and $U$
profiles appear similar to $I$.  These extent of profiles 
{\em across} the slit is naturally mixed with the
time-dependent stepping of the FIRS slit.  The $I$-like $Q$ and $U$
profiles seen in region ``A'' were observed mostly in the 9 scans
acquired between 17:46 and 17:48 UT, covering 2.7\arcsec. 
Figure~\pref{fig:more} shows
samples of these spectra.  In Appendix~\pref{sec:timevar} we argue
that the measurements are mostly of solar origin.

\figmore

The RHESSI data of Figure~\pref{fig:rhessi} 
were processed as follows.  We reconstructed CLEAN images 
\citep{Hurford+others2002} 
for each duration of a FIRS
raster step (12 s) for the 30-70 keV range of RHESSI.  The resulting
maps were then rolled by 0.2 degrees to align RHESSI to FIRS (which
was aligned to SDO/HMI, \citealp{Kleint+others2015}). 
We then
constructed a ``RHESSI raster'', simulating what it would have recorded,
had it scanned across the solar surface like FIRS. For each solar Y
coordinate of a FIRS raster step, the RHESSI reconstructed intensities 
at that time and
location were saved in a new map. Thus, only those contours 
above 30\% of the newly constructed map were plotted in Fig.~\ref{fig:rhessi},
to avoid low intensity artifacts arising from the CLEAN algorithm. This
imperfect treatment of RHESSI data serves to try to illustrate the locations
of the HXR emission from a sequence of RHESSI CLEAN images.

\subsection{A simple model for the \ion{He}{1}-emitting plasma}
\label{sec:intensities}

Figure~\pref{fig:lineplot} shows profiles from two representative
points.  Dashed lines show data from  outside of
the bright ribbon, solid lines those from within the ribbon 
itself.  Let us consider the ribbon emission during the
impulsive phase.  Under optically thin conditions, the intensity ratio
between the blended 1083.0 and single 1082.9 components will be the
ratio of the statistical weights of the upper levels, i.e., $(5+3):1=8:1$,
whether in emission or absorption.

\lineplot

Care must be taken in analyzing the intensity profiles, a na\"ive
inspection of the profile indicates the peak near 1082.9 contains
roughly 1/8 of the energy of the 1083.0 peak.  
But the 1082.9 and 1083.0 nm components have very different
intensity profiles, the 1083.0 component having an extended red wing. If we
insist that the emission is optically thin we must conclude that the 
wavelength-integrated ratio is far less than the peak ratio of 1/8 
because the 1082.9 component will have its own red wing
emission sitting under the 1083.0 nm component.  We therefore conclude
that the intensity during the impulsive phase is {\em incompatible with the
optically thin ratio of 8:1.  The emitting plasma is therefore optically thick
in at least the 1083 nm components.}

We estimate the optical depth of the \ion{He}{1} emitting plasma using
a parameterized non-LTE slab model. Such a simple model is appropriate
for the problem at hand, a slab geometry being similar to the kind of
structures in the middle-upper chromosphere found in beam-heated
hydrodynamic flare models \citep[e.g.,][]{Allred+others2005}.  The
flare ribbons seen in \hemults{} extend over areas of $\approx 30$
square arcseconds, with horizontal scales of a few thousand km, much
larger than the 100km thickness of slabs in such flare models.
Instead of solving multi-level non-LTE rate equations we adopt a line
source
function $S_L$ for a two-level atom and we iterate between this linear
equation (\pref{eq:s}) 
and the transfer equation (\pref{eq:j}):  
\be{s} 
   S_L = \epsilon B + (1-\epsilon) \bar J. 
\ee
\be{j}
\bar J = \Lambda [S_L].
\ee
Here, both $B$ and $\epsilon$ are assumed constant for each 
slab, $B$ is the 
Planck function (temperature) within the slab,
$\epsilon = C_{21}/(C_{21}+A_{21})$ where
$C$ and $A$ are collisional and spontaneous decay transition
probabilities between the two levels.   The variable 
$\bar J$ is the
mean intensity averaged over the absorption line profile.
Equation (\pref{eq:j}) gives the solution to the transfer equation for a given
$S_L$, computed using 
a lower boundary representing the photosphere as
a Planck function with temperature 
5700 K, and an the upper boundary having 
no incoming radiation.  
(The``$\Lambda$''-operator (\ref{eq:j}) integrates over depth, over  
three slant angles to the slab normal, and over 
frequencies where the line has some opacity).  
Lambda iterations
were used to bring these equations to convergence 
using a Feautrier solver, 
because the optical depths derived were modest, $\lta 20$.  

Although the 1083nm multiplet involves transitions between 
four levels, a two-level atom
suffices under conditions of source function
equality in multiplets \citep{Mihalas1978}.  Using a genetic algorithm
(GA) we searched the parameter space for values of the slab optical
depth, the non-LTE parameter $\epsilon$  and the Planck function (i.e.,
temperature) of the slab plasma.  At the same time we searched for two
unresolved plasma ``elements'' (to avoid confusion with different
line ``components'' we call these ``elements'') giving line profiles
inside the slab, keeping the opacity ratios fixed to the atomic values
for the three lines.  These elements are defined by their relative
opacity, the line widths and Doppler shifts.  It is clear from the emergent 
$I$
profiles (Figure~\pref{fig:lineplot}) that at least two elements are
needed, one to account for the relatively narrow 1082.9 nm peaks and
another for the extended red wing and broader peak of the 1083.0 nm
components.

Results of the non-LTE intensity calculations are shown in
Figure~\pref{fig:ifit}.  The top left panel shows the intensity data
from the center of box ``A'' in Figure~\pref{fig:comparef}.
The dashed line shows the optimal fit from the slab calculations, and
the dot-dashed line shows the run of opacity with wavelength from the
two elements combined in the \hemults{}.  The line center optical depth
of the fitted slab is 15, $\epsilon=0.011$ and the Planck function is 10
times the Planck function in the photosphere.  All intensities shown
in the figure are divided by the photospheric Planck function.  
The role of the finite optical depths is seen in the GA
solution (dashed line) that is far broader than the elementary opacity
itself. The solution is also slightly self-reversed at the peak
intensity.  The GA's solution is very different from the dot-dashed
opacity which would represent the solution under optically thin
conditions: the ratio of the 1082.9 to 1083.0 components being closer
to observations in the case of finite optical depth.

\figifit

\subsection{Origin of the Impulsive Phase Linear Polarization}

To produce emission line intensity from atoms in a solar plasma, the
environment must generate a finite {\em population} in a given level,
for example by impact with electrons or by irradiation.  To produce
polarization, the environment must introduce anisotropy to break
symmetries often present under natural conditions (such as in thermal
equilibrium, or LTE in cylindrically-symmetric geometry).  In the
spherical tensor basis \citep{Fano1957}, spectral line polarization
requires non-zero atomic {\em alignment} and/or atomic {\em
  orientation} \citep{Landi+Landolfi2004} of atomic levels.  Atomic
population, alignment and orientation lead to intensity, linear and
circular polarization in spectral lines respectively.  For \HeI{} 
three long-lived atomic levels need be considered as reservoirs from
which significant population, alignment and orientation can be
produced (Figure~\pref{fig:td}).  Note that the linear polarization
that we seek to explain has a magnitude that is of the order of the
circular polarization and very different spectral and spatial profiles
(Figure~\pref{fig:comparef}).  Thus we look to mechanisms that
generate alignment from asymmetries and ignore possible
orientation\--to\--alignment effects.

We first examine radiative anisotropies and  
discuss collisional polarization in Appendix~\pref{sec:collisions}. 
In the  natural basis of spherical tensors \citep{Fano1957}, 
the lowest order term influencing linear polarization
is the tensor component of the mean (i.e., wavelength-averaged) 
intensity $J^K_Q$ with $K=2$ and
$Q=0$ \citep[e.g.][]{Casini+Landi2008}.  The mean intensity $J^0_0$ and the 
term
$J^2_0$ are, under the cylindrical symmetry present in the slab
\citep[][equation 5.164]{Landi+Landolfi2004},
\begin{eqnarray}
J^0_0 &=& \frac{1}{2} \int_{-1}^{+1} I(\mu) d\mu\\
J^2_0 &=& \frac{1}{4\sqrt{2}} \int_{-1}^{+1} 
(3 \mu^2 -1) I(\mu) d\mu
\end{eqnarray}
where $\mu=\cos\theta$ with $\theta$ the angle of the ray to the
normal of the slab.  Generally speaking, $J^2_0$, when averaged over
the absorption line profile, is a radiative source term for linear
polarization (see equation~\pref{eqn:2level}).  The right uppermost panel
of Figure~\pref{fig:ifit} shows $J^2_0$ just above the slab (which
consists only of outward directed radiation).  The lower panels show
$J^0_0$ and $J^2_0$ as functions of wavelength across the 1083nm
region and depth in the slab.  It is clear that {\em with slab optical
  depths $>1$, radiation anisotropies in \hemults{} can readily
  approach a few percent of the mean intensities.  Also,
 an optically thick 1 dimensional slab generates $J^2_0$
values which change sign with wavelength, in particular 
between the 1082.9 and 1083.0 nm 
components}.

Given the three reservoirs of
population, two radiative processes can be important for
the \hemults{}. The first is photoionization by photons below 50.4 nm from
level 1 to 3, leaving level 3 to be impacted by electrons to produce
population in the $1s2p~^3P^o_{2,1,0}$ level via recombination and
cascades.  Level 3, with $J=1/2$, is intrinsically unpolarizable for
$^4$He isotopes.  Thus to produce any atomic alignment in a
$1s2p~^3P^o_{2,1}$ level through photoionization followed by
recombination requires an anisotropic distribution of captured
electrons (i.e., electrons $e^-(k \ell)$ with energy $k$ and angular
momentum $\ell$ cannot be drawn from an isotropic distribution).  This
origin of polarization is therefore collisional, which 
discussed separately in Appendix~\pref{sec:collisions}.

The second radiative process to consider is photo-excitation from
level 2 to $1s2p~^3P^o_{2,1,0}$, i.e. photo-excitation in the
\hemults{} itself, or to higher levels in the triplet system that by
radiative cascade transfer alignment to the $1s2p~^3P^o_{2,1}$ levels
themselves.  For our purposes it suffices to ignore cascade
contributions to alignment and orientation.  The simple non-LTE
radiation transfer slab model used above yields the needed anisotropy
of the radiation field in the \hemults{} at a level of a few percent.  But these
values are merely in the radiation field, related to but not identical to
the Stokes components that emerge from the radiating helium atoms.    
The radiative transfer determining the actual polarization in 
the three transitions is considerably more complex than the two-level atom
discussed here \citep[][section 12.4]{Casini+Landi2008}, especially because the 
$J=1$ lower level is polarizable. However, the slab and two-level 
approximation for the linear polarization yields the right order of
magnitude for the atomic polarization 
as we can see from the following.   
The simplest possible case is a two-level
atom with an unpolarizable lower level, the solutions to which
yield 
\citep[][equations 10.50 and 10.51]{Landi+Landolfi2004}:
\begin{equation} \label{eqn:2level}
\sigma_0^2= { {D_{J J_0} \bar J_0^2 } 
    \over 
{{\epsilon B_\nu(T) + \bar J_0^0}}}
\, \times \,
{{1+\epsilon}\over{1+\epsilon+\delta_u^{(2)}}},
\end{equation}
\noindent where $\sigma_0^2$ is the ratio of atomic alignment to
atomic population for the upper level, 
 $D_{J J_0} $ (of order 1) 
depends only on the quantum numbers of
the two-level atom, $\epsilon$ is the ratio of the downward
(super-elastic) collision rate to the Einstein A-coefficient,
$B_\nu(T)$ is the Planck function at the coronal temperature $T$, and
$\delta_u^{(2)}$ is the depolarizing collisional rate of the upper
level.  $\sigma_0^2 $ is the leading term in contribution to the
fraction of linearly polarized light, it is linearly proportional to
the frequency averaged value, $\bar J^2_0$. With small values of
$\epsilon$, $\epsilon B_\nu(T)/J_0^0$ and 
$\delta_u^{(2)}$, $\sigma_0^2$ becomes $ \approx \bar J^2_0/ \bar J^0_0$.
In words, the ratio of alignment to population is 
$ \approx \bar J^2_0/ \bar J^0_0$, so that the observed 
ratio $P/I$ is also $ \approx \bar J^2_0/ \bar J^0_0$. 

By examining various slab models, we find that the primary source of
atomic polarization in the \hemults{} is from non-zero values of $
J^2_0/ J^0_0$ generated naturally {\em scattering in slabs which have
  an optical depth $\tau_0$ in the core of the strongest lines of at
  least 1}.  At smaller optical depths the radiation field is modified
weakly by the helium lines and we see almost pure photospheric
continuum, which far from the solar limb is very weakly polarized ($P/I_C<
0.0001$, \citealp{Stenflo+others1997}).  The best fits to the observed
intensity profiles, such as that shown in Figure~\pref{fig:ifit},
produce linear polarization of a few percent when slabs have 1083.0 nm 
optical
depths $\approx 10^1$.  In such models 
$ J^2_0/ J^0_0$ changes sign from $<0$ at
wavelengths where the slab is optically thin, to $>0$ for wavelengths
where it is optically thick.  Since the same sign of $J^2_0/ J^0_0$
generates polarization in the 1082.9 nm component that is of opposite
sign to those in the 1083.0 components 
\citep[table 10.3 of][]{Landi+Landolfi2004},
the effect is to generate polarization of {\em the same sign for all
  three components} under the conditions shown in the right panels of
Figure~\pref{fig:ifit}.
If the optical depth
varies from place to place in the flare ribbons, the linear
polarization will vary accordingly.  In particular, optically thin
slabs will show very little polarization even though they might even have
comparable intensities.  This picture appears broadly compatible with
the diverse observations reported here and in 
\citet{Kuckein+others2015a,Kuckein+others2015b}, which have $P/I\approx 0.1\%$.

\revise{From the definition of Stokes vectors, the electric vector of
  the polarized radiation lies along the direction in which, when
  rotated to a frame rotated on the plane-of-sky, Stokes $U$ becomes
  zero.  This places the electric vector along the direction
  $\frac{1}{2} \arctan~U/Q$, which for the 1083.0 nm component 
  is $\approx 15^\circ$ wrt the E-W direction.
  Observed $U/Q$ ratios seen during the flare are close to 1/2.  Since
  the polarization has a radiative origin, and the magnetic field
  strengths of several hundred G (paper I) greatly exceed the
  ``Hanle'' field strength which for 1083nm is close to 1G
  \citep{Casini+Landi2008}, the electric vector under this condition
  has a $90^\circ$ ambiguity.  Sun center is at $30^\circ$ to the E-W
  direction, The \ion{Si}{1} Zeeman patterns from paper I place the
  underlying photospheric field close to the Sun center direction,
  with a statistical fluctuation of around $40^\circ$ in the pre-flare
  footpoints.  Our conclusion is that the electric field vector
  probably lies parallel to the magnetic field in the plane of the
  sky, not perpendicular to it. This simply means that the atomic
  alignment for the 1083nm component 
  is positive, which is entirely consistent with the
  calculations presented above (Figure~\pref{fig:ifit}).

\cite{Stepan+Heinzel2013} calculated polarization in a stratified
atmosphere assuming a two dimensional geometry to study effects of
asymmetries in radiation at the {\em edges} of flare ribbons.  Our
slab calculations are consistent with theirs away from the regions of
strong horizontal gradients.  Positive alignments are found within
ribbons which have a physical width $\gg \lambda$ where $\lambda$ is a
local photon mean free path, and where transport is essentially only
vertical.  We note that $\lambda$ is expected to be on the order
of a pressure scale height, perhaps $10^2$ km, far smaller than the ribbon's
width of 3 Mm or more (Figure 1).
Our work differs from \cite{Stepan+Heinzel2013} mainly through our
assumption of a slab, and by the fact that we look at all three
components of a multiplet sampling very different optical depths.
The change in sign of alignment between the 1083 and 1082.9 nm 
components implied by Figure~6 is an entirely new and unexpected
result.

}

Finally, in Appendix~\pref{sec:collisions} we review collisional
origins of the \hemults{}, assuming that energetic
particles responsible for HXR  emission are anisotropic and that they 
directly generate atomic polarization \citep[e.g.][]{Henoux+Chambe1990}. 
Collisions cannot by themselves produce 
polarization in 1082.9~nm, this requires the same lower-level
polarization via radiation transfer outlined above.  They also cannot 
readily account for the spatio-temporal differences between the HXR
and helium polarization data shown in Figure~\pref{fig:rhessi}. We conclude,
based upon collisional relaxation times, that any direcy contribution to 
the polarization from anisotropic particles most likely would have to
arise from protons near 1 MeV.  Given the success of the radiative models
above though, there is no reason to invoke collisions to explain the data. 

\section{Discussion and Conclusions}
\label{sec:disc}

We have analyzed data for the \hemults{} during the X1 class flare
SOL2014-03-29T17:48.  We found that the \HeI{} profiles are formed in
an optically thick layer lying above the solar photosphere.  The
peculiar $Q$ and $U$ Stokes profiles are dominated 
by a real solar signal.  The most probable origin of
the observed linear polarization during the impulsive phase is {\em
anisotropic photon scattering in a slab of optical depth of order 10},
somewhere above the photosphere, with no need to
invoke collisionally-induced anisotropy.  The high energy electrons
required to explain the HXR emission seen with RHESSI serve
not as a direct cause of atomic polarization through collisions from
an anisotropic electron population, but merely as a source of energy
that, as it thermalizes, generates thermal emission from the slab.

A remarkable and unanticipated success of this simple model is the
fact that it predicts that the 1082.9 nm and 1083.0 nm components will
have the same sign, as observed.  Under previously discovered
conditions in prominences and filaments, lower level polarization in
the 1082.9 nm component leads to linear polarization of the opposite
sign to the polarization seen in the 1083.0 components
\citep{Trujillo+others2002}.  However, here the observations clearly
reveal the same sign of polarization in the entire multiplet.  The
explanation lies in the change of sign of the anisotropy, $J^2_0$
between these two components (top right panel of 
Figure~\pref{fig:ifit}). This sign change
arises naturally from the slab geometry under the presence of
significant thermal emission from the slab itself, and when the
strongest 1083.0~nm components have optical depths of order ten, with
the 1082.9~nm component having optical depth of order one.

The impulsive phase of the flare thus appears to heat the upper
chromosphere sufficiently to produce emission in the entire \hemults{}
and at the same time the fit to the intensity spectrum leads naturally
to conditions where the three components must have the same sign in
linear polarization.  This is also, perhaps, a hint that the far red
wing of the 1083.0 nm component shows a negative polarization (see the
lower middle panel of Figure~\pref{fig:lineplot}), but we remind the
reader that the polarization fringes are not negligible in comparison
with these weak signals (see the diffuse regular vertical stripes in the
upper middle panel).  In Appendix \pref{sec:collisions} we argue that
collisional impact polarization is unlikely to produce the linear polarization
observed.  In fact, the 1082.9~nm transition cannot be polarized
without lower level polarization.  It is difficult to see how this can 
be produced effectively by collisions alone.

To test this conclusion further we suggest the following.  First, we
must obtain the highest sensitivity data possible for additional
impulsive phases.  Second, we must have a strategy to reduce instrumental artifacts such as  fringes.
If we can obtain reliable data with a sensitivity of $P/I \lta
10^{-3}$ with negligible residual systematic errors, then we can seek
the presence of the predicted change of sign in linear polarization
between the 1083.0 and 1082.9 nm components that must arise from the
multi-level transfer in slabs of optical depths greater than one
\citep[see figure 12.16 of][]{Casini+Landi2008}.  Third, we must
develop a radiative transfer program to handle the non-LTE radiation
fields and density matrices for the radiating atom at least in slabs.
Section 12.4.2 of \citet{Casini+Landi2008} presents an initial model
based upon two-level atoms, with lower (upper) levels with $J=$ 0 (1)
and 1 (0) respectively.  The latter case exhibits the lower level
polarization or ``dichroic polarization'' observed in a filament of
finite optical depth \citep{Trujillo+others2002}, with the tell-tale
change in sign of polarization between the 1083.0 and 1082.9 nm
components that characterize transfer through a slab of optical depth
$\gta 1$.  Unfortunately, the residual fringes and crosstalk present
in our data confuse this critical aspect of our particular data.

If we are later proven incorrect, and collisions are important for
generation of \hemults{} polarization, there will be additional
implications.  Conflicting constraints on the energy needed to produce 
polarizing electron-
helium collisons suggests 
that instead {\em protons} might conceivably cause some  polarization seen at the
footpoints of this X1 flare.  In this picture, the polarization is
generated by an anisotropic distribution of protons of $\approx 1$
MeV impacting the top of the chromosphere \citealp[see
  also][]{Henoux+others1990}.  

Finally, the diverse results in the literature regarding the H$\alpha$
\citep{Henoux+Chambe1990,Henoux+others1990,Hanaoka2003,Bianda+others2005}
and now \hemults{} polarization (compare our measurements with those
of \citealp{Kuckein+others2015a,Kuckein+others2015b}) can readily be
explained once the optical depth of a slab is recognized as a primary
determinant of the outgoing polarized light.  This is a quite
different model from an earlier picture where anisotropically
distributed particles are responsible for linearly polarized light from
chromospheric spectral lines \citep{Henoux+others1990}.

\acknowledgments Data in this publication were obtained with the
facilities of the National Solar Observatory, which is operated by the
Association of Universities for Research in Astronomy, Inc. (AURA),
under cooperative agreement with the National Science Foundation.
Part of this work was carried out under NASA grants NNX13AI63G
and NNX14AQ31G.  We
thank the observers at the DST for their help and Javier Trujillo
Bueno, Fatima Rubio de Costa 
and Marina Battaglia for helpful discussions.  Christian Beck 
was extremely helpful in demonstrating usage of the 
NSO data reduction package. We acknowledge
other helpful discussions that took part at ISSI.

\appendix
\section{A. Time variations, flares, and crosstalk}
\label{sec:timevar}

Here we review
origins of errors in the measurement process that arise because light
coming into spectropolarimeters changes with time.  One complete
measurement of all four Stokes parameters $IQUV$ requires at least
four measurements of the type
\be{state}
P_i = I + a_i Q + b_i U + c_i V,
\ee
where $P_i$ is termed the polarization state that is measured 
between times $t_i$ and $t_i + \Delta t$.  
In FIRS the $i=1\ldots4$ measurements are sequential in time, one
complete cycle taking 1.3 sec for the data acquired here.  To retrieve 
the quantities $I,Q,U,V$ we must make at least four such measurements 
each with known values of $a_i,b_i,c_i$ such that equation~(\pref{eq:state})
can be inverted.  The modulation
cycle used in the observations can be written 
\be{mod}
{\bf P} = {\bf M~S} 
\ee
where ${\bf S} = (I,Q,U,V)^T$ The 
modulation matrix ${\bf M}$ used in the observations reported here is
\newcommand{\Sr}{\sqrt{3}}
\be{mod1}
{\bf M} = \frac{1}{2} 
\begin{pmatrix}
  1 & -\Sr & -\Sr & \Sr\\
  1 & -\Sr &  \Sr &-\Sr\\
  1 &  \Sr & -\Sr &-\Sr\\
  1 &  \Sr &  \Sr & \Sr
\end{pmatrix}
\ee in which $U$ (the third column) is clearly modulated at twice the
rate of $Q,V$.  Since usually $|I| \gg |Q|$, $|U|$ and $|V|$, two
beams of light are measured for each accumulation state $i$ with the
second beam measuring $I_i = I - a_i Q - b_i U - c_i V$.  Subtracting
the two beams removes, to the level at which the beam intensities can
be calibrated, $I\rightarrow QUV$ crosstalk. Taking care of the
alignment and relative gains of the two beams (dark current and flat
field calibrations), the inverse matrix ${\bf M}^{-1}$ applied to
${\bf P}$ yields $IQUV$ {\em at the entrance to the polarimeter}
from each complete measurement cycle.  To convert these measurements to the solar Stokes parameters requires inversion of a calibration matrix, taking account of the 
modification of the solar Stokes parameters by the telescope.

Under constant illumination, the above scheme will retrieve solar Stokes
profiles to within uncertainties defined by the flat field, gain
calibrations, and photon counting noise.  However, if the incoming
light varies on times comparable to of shorter than the accumulation
cycle ($4 \Delta t)$, systematic errors are induced.  The most
important time variations are probably due to atmospheric seeing, which
introduces random errors proportional to the spatial gradients of the
light across the solar surface
\citep{Lites1987,Judge+others2004,Casini+deWijn+Judge2012}.  The
spatial gradients are in general wavelength-dependent.  If, for
instance, the solar photosphere has more contrast in a line than
in the neighboring continuum, seeing-induced noise will be larger in the
spectral lines.  When, as is usual, Stokes $I$ is larger than $QUV$, 
variations in $I$ are translated, by this source of 
measurement error, into $Q,U$
and $V$.  Such sources of error are termed ``crosstalk''.

In studying flares we must also be aware of time variations occurring
in the solar plasma itself \citep{Judge+others2004}.  In Paper~I we
found 4\% changes in the IR continuum intensity in the flare
footpoint, but the core intensities of the \ion{Si}{1} and \ion{He}{1}
lines change by factors of 2 or more (see Figures~\pref{fig:comparef}
and \pref{fig:more}), most likely on time scales of  seconds given the 
RHESSI HXR observations.    Under
normal conditions, the \ion{He}{1} 1083 nm multiplet forms at the base
of the corona in tenuous plasma that also contributes to UV radiation.
UV line and continuum brightnesses change by orders of magnitude
during flares \citep[e.g.][]{Brekke+others1996}.  Thus we cannot
discount the idea that, on time scales of seconds or less, the
\ion{He}{1} line's intensity might change considerably.  In that case,
we would have to consider the possibility that the $QUV$ profiles
contain significant crosstalk from $I$.

The $Q,U$ signals of a few percent exceed by at least 20$\sigma$ the
random noise levels of $\approx 8\times10^{-4}I_c$ (Paper~I).
However, {\em crosstalk} can induce much larger spurious polarization
signals \citep[e.g.][]{Lites1987}.  Crosstalk originates because the
measurement of Stokes profiles takes time, during which at least four
independent integrations in four different states of polarization
must be made.  In FIRS this is done
sequentially in time, requiring 1.3 s for a complete cycle.  If the
Sun, atmospheric seeing/clouds or observing system introduces
variations on time scales less than this cycle time, measured
quantities contain mixtures of the average solar conditions (the
``real'' Stokes vectors sought) and other variations.  When
demodulated to produce the measured Stokes profiles, the profiles can
contain systematic errors.  

At the DST, light level and scintillation (in arc seconds) are
monitored and the data are reported to the instruments.  From the FIRS
headers, the light levels were constant to within 1.6\%.
Scintillation was $1.2\arcsec \pm 0.35\arcsec$ (mean and rms) 
during the impulsive phase, slightly above the mean of 
$0.88\arcsec \pm 0.44$ of the entire scan.  At wavelengths near
\hemults{} the image intensity contrasts are also slightly higher
during the flare.  Therefore seeing-induced $I\rightarrow QUV$ crosstalk cannot
be immediately ruled out.

In spite of these issues we argue that the strong $QU$ profiles
observed during the flare are of solar origin, based upon several
properties of the data themselves.  

\begin{itemize}

\item Similar profiles are observed at all times throughout box ``A''
(Figure~\pref{fig:more}), under conditions of different seeing. Those shown
in Figure~\pref{fig:comparef} are typical profiles. 

\item The impulsive phase $QU$ profiles, while similar to Stokes $I$,
  are however not identical.  Seeing-induced polarization is to a
  first approximation proportional to the rms seeing and spatial
  gradients in the solar Stokes profiles \citep{Lites1987}.  If
  crosstalk from $I$ were dominating the polarization signals,
  $I$-like profiles would appear in all of $QUV$, for example as
  $U_{\rm measured} = U_{\rm solar} + c I_{\rm solar}$. To estimate
  an upper limit for 
  $c$ we assume that all of the ``sharp peak'' in Stokes $U$ near
  1082.9 nm (i.e. the ``blue'' component) is due to crosstalk, with the measured values
  $c \approx 0.005$.  Figure~\pref{fig:comparef} however shows that the
  particular profile plotted as a line in \pref{fig:lineplot} is not
  representative of other areas within box ``A''.  Indeed, both $Q$
  and $U$ are qualitatively quite different in the lower part of box
  ``A'' from $I$.  Thus the measured $QU$ profiles, although they can
  contain some crosstalk from $I$, cannot be dominated by it.

\item Independent reduction codes (those of S. Jaeggli, T. Schad and
  C. Beck) yield very similar results. 

\end{itemize}

\figseeing

The observed behavior indicates that crosstalk in the \hemults{} 
must either be {\em substantially below the $c=0.005$ level, or it
  must vary with the position along the slit}.  The data were acquired
with the Dunn Solar Telescope's (DST) adaptive optics system (AO)
switched on. The AO system corrects for image motions mostly in an
isoplanatic patch of several seconds of arc near the center of the
field.  The center of the AO field was close to the pore at
$X=525\arcsec$ and $Y=270\arcsec$ \citep[][]{Kleint+others2015}. The AO
correction was of variable quality across these areas, judging from the 
G-band images acquired simultaneously, close-ups of which are shown in
Figure~\pref{fig:seeing}.   
All things considered, we conclude that seeing-induced crosstalk 
does not dominate the measured $QUV$ signals, except perhaps for the 1082.9 nm
components of $QU$ which are weaker and more susceptible to 
remaining systematic errors.

\section{B. Collisional origins of linear polarization}
\label{sec:collisions}

Anisotropic particle distributions are known to be associated with
flares from {\em in situ} measurements in interplanetary space, 
and are a natural consequence of most flare particle-acceleration mechanisms
\citep[e.g.,][]{Simnett1995}.  Atomic 
polarization is readily induced by 
collisions of radiating atoms with 
anisotropic particles 
\citep[e.g.,][]{Henoux+Chambe1990}, analogously to radiation anisotropy. 
Given the coincidence of at least the strongest $Q$ and $U$ profiles with 
the HXR footpoint emission (Figure~\pref{fig:rhessi}), we cannot 
immediately discount particle 
collisions as a contributing 
source for the observed \hemults{} 
linear polarization.  Therefore here we review the basic processes 
by which particles may generate polarization in the \hemults{}. 

The dominant particles in the Sun's chromosphere are 
electrons, protons and hydrogen atoms.  We consider terms in the Boltzmann
transport equation that drive (i.e., force away from equilibrium) 
and relax (to equilibrium and, in particular, isotropy) 
the distributions of hydrogen atoms, protons, and electrons.
Neutral hydrogen atoms interact only weakly with electric and magnetic
fields, and the self-collisional relaxation time is short, $\tau_{\rm
  H-H} \approx \left (\pi a_0^2 n_H v_{\rm thermal}\right )^{-1}
\approx 10^{-3} $ sec for $n_H = 10^{11}$ cm$^{-3}$ (the hydrogen
density varies from $10^{11}$ to 10$^{15}$ cm$^{-3}$ across the
ambient chromosphere).  Note that $\tau_{\rm H-H} \propto \en^{-1/2}$
where $\en$ is the kinetic energy in eV of the colliding particles
before impact.  This weak driving and strong relaxation means that
hydrogen atoms are expected to be isotropic to a high degree of
approximation.  Impact with hydrogen can serve only to depolarize
existing atomic polarization in \ion{He}{1}.

Both electrons and protons are accelerated by electric fields,
deterministic or stochastic, as such they are expected to show
departures from isotropy during flares. The determination of such
fields remains one of the goals of flare research
\citep{Fletcher+others2011}. Therefore we first look at relaxation
times.  Relaxation self-collision times for protons and electrons are
dominated by the accumulation of small angle scatterings under the
long-range Coulomb potential, and they have a different behavior with
impact energy \citep{Braginskii1965}:
\begin{eqnarray}
\tau_{\rm e-e} &\approx& \frac {3.5\times 10^4 \ \en^{3/2}}{(\lambda/10) n_e} \\
\tau_{\rm p-p} &\approx& \frac {2\times 10^6 \ \en^{3/2}}{(\lambda/10) n_p}
\end{eqnarray}
where $\lambda$ is the Coulomb logarithm (between 10 and 35 for most
plasma in the solar atmosphere), $n_{e,p}$ are ambient electron and
proton densities.  With $n_e = n_p \gta 10^{11}$ cm$^{-3}$ throughout
the chromosphere we find collision times of $\lta 10^{-7}\en^{3/2}$
and $\lta 10^{-5}\en^{3/2}$ sec for electrons and protons
respectively.  To retain an anisotropic distribution function, the
particles that impact helium atoms must survive collisional relaxation
on their journey from the source of acceleration into plasmas
emitting the helium spectrum. This constraint would seem to rule out
particles at low energies.  
The accumulations of many Coulomb collisions leads to a stopping or
penetration depth in column mass units, such that, for electrons
\be{pen}
m_p = 1.4\times10^{-5} \times \left ( \frac{\enk}{20} \right )^2 ~{\rm g~cm^{-2}}
\ee
This leads to a constraint ``C1'', a lower limit 
on the energy of particles that can 
travel through the chromosphere.
In the standard model, the pre-flare chromosphere is at column masses of $\gta
10^{-5}$ g~cm$^{-2}$, so that only electrons of energy $\gta 20$ keV
or protons of energy $\gta 1$ MeV can enter the chromosphere from
above and retain some anisotropy.

To generate significant polarization in neutral atoms by electron or
proton collisions requires an anisotropic particle distribution at
energies $\eno$ where the absolute cross sections $\sigma(\eno)$ 
are sufficiently large
(constraint ``C2''), and where the cross sections for the
magnetic sub-states are also sufficiently different (constraint ``C3'')
such that sub-state populations become unequal.
For inelastic collisions (ionization, 
recombination, collisional excitation and de-excitation), C2 and
C3 occur together only under rather restrictive conditions. Further, 
these conditions are generally in conflict with C1. 

For electron-ion collisions, consider the  recombination process
\be{rec}
1s~^2S_{1/2} + e^- (\eno,\  \ell=1)   \rightarrow 1s2p~^3P^o_{2,1,0} + h\nu
\ee
and similar processes for recombination to higher levels ($1s nl$
configurations, $n > 2$) which cascade down the triplet system of
helium.  Here, $\eno$ is the average energy of electrons with angular
momentum $\ell$.  The cross sections for this process are relatively
small (compared with other collisional processes), yielding rates of
recombination per He$^+$ nucleus that are $\approx 10^{-13} n_e
\en^{-1/2}$ s$^{-1}$, using hydrogenic coefficients from \citep[][\S
  38]{Allen1973}.  The conflict of constraint C1 with C2 is evident.
   {\em Recombination is improbable at the energies where
     electrons survive long enough to maintain anisotropic
     distribution functions. } 

For collisional excitation, we use the theory of impact polarization
by \citet{Percival+Seaton1958} to estimate differential cross sections
for collisions between different sub-levels (with magnetic quantum
number $M$).  In the absence of electron exchange (which is usually
negligible for electric dipole transitions, varying as
$\varepsilon^{-3}$ \citep{Seaton1962a,Burgess+Tully1992} and ignored
in the Percival-Seaton theory), the collision cross sections are
determined by the impacting particle's path and charge.  For neutral
targets, the paths of electrons and protons are essentially the same
(at least at  energies greatly in excess of the 1-10 eV bound
energies), being represented by plane waves and an outgoing spherical
wave.  Thus the cross sections for protons should be very close to
those computed using the theory for electrons.  
The only long-lived level from which collisions with energetic
particles 
might directly excite and polarize the \hemults{} is $1s2s~^3S_{1}$,
the other long lived \ion{He}{1} levels requiring 
spin changing transitions with small cross sections. 
Therefore, we consider the process
\be{col}
1s2s~^3S_{1} + e^- (\eno,  \ell) \rightarrow 1s2p~^3P^o_{2,1,0} + 
e^- (\eno',\  \ell'=\ell\pm1)
\ee
Collisions to the
$1s2p~^3P^o_{2,1,0} $ upper levels occur via electric dipole
transitions, for which the collision cross section varies as
\citep{Seaton1962a}
\be{e1}
\sigma \propto \frac{\ln \eno }{\eno}
\ee
Constraint C2 is in conflict with C1 again: At very high energies the
probability of exciting the $1s2p~^3P^o_{2,1,0} $ upper levels drop
rapidly with increasing energy.  The differential cross sections of
\citet{Percival+Seaton1958} at very high energies can be estimated
using the Born approximation (Section 6 of
\citealp{Percival+Seaton1958}).  The cross sections maintain the same
differential atomic polarization (i.e. constraint C3).  We conclude
for the \hemults{} that there is a ``sweet spot'' in energy such that
the cross sections for generating large populations are large (C2,
low-ish energy) but that the energies are sufficiently high (C1) that the
impacting electrons or protons penetrate the chromosphere and remain
anisotropic.

\percival

Figure~\pref{fig:percival} shows the cross sections computed using the
theory of \citet{Percival+Seaton1958} using cross section data
computed from \citet{Flannery+McCann1975}.  The right hand panel shows
the net polarization induced by collisions with electrons and protons,
parallel and perpendicular to a collimated beam of such particles.
For electrons, constraints C1 forces us to adopt a population of
electrons with energies $\gta 20$ keV.  But at such energies the cross
sections become very small (as seen through extrapolation of the 
left panel of Figure~\pref{fig:percival}
using Equation~(\pref{eq:e1})).  The left hand panel of the figure shows
that a population of protons above 1 MeV (C1) 
might satisfy constraints C2
and C3.  In words, MeV protons might just account for the observed
polarization in the 1083.0 nm transitions.  

Note however that this theory produces polarization {\em only} in the 
two 1083.0~nm transitions whose upper levels are polarizable 
($J=1$ and $2$). 
In the main text we emphasize the need for {\em lower level}
polarization to produce polarization in the 1082.9~nm line which has
an unpolarizable ($J=0$) upper level.  To populate this level at all,
let alone polarize it, one requires an electron exchange transition
from level (1) or recombination from level (3).  In either case the
probability of generating a polarized lower level, independent of the
radiation field, is small.  Thus we conclude that collisions are not
the primary source of the observed polarization. 

\vskip 24pt


\end{document}